\documentstyle[pra,aps,epsfig,graphics]{revtex}
\begin{document}
\title{Quantum Teleportation and Bell's Inequality Using
Single-Particle Entanglement}
\author{Hai-Woong Lee\footnote{E-mail address :
hwlee@laputa.kaist.ac.kr} and Jaewan Kim\footnote{On leave
from Samsung Advanced Institute
of Technology}}
\address{Department of Physics,
         Korea Advanced Institute of Science and Technology,
         Taejon 305-701, Korea}
\maketitle
\begin{abstract}
A single-particle entangled state can be generated by illuminating
a beam splitter with a single photon. Quantum teleportation
utilizing such a single-particle entangled state can be
successfully achieved with a simple setup consisting only of
linear optical devices such as beam splitters and phase shifters.
Application of the locality assumption to a single-particle
entangled state leads to Bell's inequality, a violation of which
signifies the nonlocal nature of a single particle.
\end{abstract}
\vspace{1cm} PACS number(s): 03.67.-a, 03.65.Bz, 42.50.-p

\section{Introduction}

It has long been realized that the striking nonclassical nature of
entanglement lies at the heart of the study of fundamental issues
in quantum mechanics, as witnessed by the Einstein-Podolsky-Rosen
paper\cite{PR47_777}, Bell's theorem\cite{PLIC1_195} and its
subsequent experimental verifications\cite{RPP41_1881,PRL49_91}.
The recent surge of interest and progress in quantum information
theory allows one to take a more positive view of entanglement and
regard it as an essential resource for many ingenious applications
such as quantum teleportation\cite{PRL70_1895,N390_575} and
quantum cryptography\cite{PRL67_661}. These applications rely on
the ability to engineer and manipulate entangled states in a
controlled way. So far, the generation and manipulation of
entangled states have been demonstrated with photon pairs produced
in optical processes such as parametric
downconversion\cite{N390_575,PRL75_4337}, with ions in an ion
trap\cite{PRL74_4091}, and with atoms in cavity QED
experiments\cite{PRL79_1}. All these experiments use as a source
of entanglement two or more spatially separated particles
(photons, ions or atoms) possessing correlated properties.

In this work we consider entanglement produced with a single
particle(``single-particle entanglement") and explore its
usefulness. As a prototype of a single-particle entangled state,
we take an output state emerging from a lossless 50/50 beam
splitter irradiated by a single photon. Here the one photon state
and the vacuum state can be regarded to represent the logical
states 1 and 0 of the qubit. Single photons have already been
considered as a unit to carry logical states of the qubit in a
proposal to construct a quantum optical model of the Fredkin
gate\cite{PRL89_2124}. Recently, it has been proposed that the
single-photon entangled state be used to create macroscopic
entangled field states\cite{PRA62_012102}.

The main propose of this work is two-fold. First, we wish to
present a scheme for quantum teleportation based on the
single-photon entangled state. A characteristic feature of this
scheme is that it requires only linear optical devices such as
beam splitters and phase shifters and thus provides a way of
achieving all linear optical teleportation along the line
suggested by Cerf et al\cite{PRA57_R1477}. Second, we wish to
derive a single-particle version of Bell's inequality that is
applied to an interference pattern produced by single particles. A
violation of this inequality establishes the nonlocal nature of a
system described by a single-particle entangled state.

\section{Single-particle entanglement}

Let us consider a single photon incident on a lossless symmetric
50/50 beam splitter equipped with a pair of $-\frac{\pi}{2}$ phase
shifters, as depicted in Fig.\ref{beam}. Denoting the two input
ports of the beam splitter by I and J and the output ports by A
and B, and assuming that the photon enters the beam splitter
through the input port I, the input state can be written as
$|1\rangle_I|0\rangle_J$, where $|1\rangle$ and $|0\rangle$ are
the one photon state and the vacuum state, respectively, and the
subscripts I and J refer to the modes of photon entering the beam
splitter through the input ports I and J, respectively. The output
state emerging from the beam splitter is then given by

\begin{equation}
|\Psi\rangle=\frac{1}{\sqrt{2}}(|1\rangle_A|0\rangle_B
+|0\rangle_A|1\rangle_B) \label{input}
\end{equation}
where subscripts A and B refer to the modes of photon exiting the
beam splitter through the output ports A and B, respectively. The
state given by Eq.\ (\ref{input}) represents a single-photon
entangled state. We note that the output state is obtained in the
symmetric combination as given by Eq.\ (\ref{input}), because the
phase shifter at the output port A acts to offset the phase
difference of $\frac{\pi}{2}$ between the reflected and
transmitted waves\cite{OC64_485} (we assume throughout this work
that the reflected wave leads the transmitted wave by
$\frac{\pi}{2}$ in phase). The phase shifter at the input port J
does not play any role in this case because only vacuum is present
at this port.

\section{Quantum teleportation}

We are now ready to describe a teleportation scheme that makes use
of single-particle entanglement. As in the standard teleportation
scheme\cite{PRL70_1895,N390_575}, this scheme consists of three
distinct parts as shown in Fig.\ \ref{tele}; the source station
that generates a single-photon entangled state, Alice's station
where a Bell measurement is performed and its result is sent away
through classical communication channels, and Bob's station where
the signal from Alice is read through classical communication
channels and a suitable unitary transformation is performed.
Details of the teleportation procedure described below follow
closely the original proposal\cite{PRL70_1895}.

The source station consisting of the same setup as in Fig.\
\ref{beam} generates a single-photon entangled state in the form
of Eq.\ (\ref{input}). The reflected wave A of the entangled state
is sent to Alice and the transmitted wave B to Bob. At Alice's
station this reflected wave A of the entangled state is combined
via a lossless symmetric 50/50 beam splitter with a pair of
$-\frac{\pi}{2}$ phase shifters to a wave C which is in an unknown
superposition of a one photon state and a vacuum state,
$a|1\rangle_C+b|0\rangle_C$, where $|a|^2+|b|^2=1$. This state of
unknown superposition is the state that Alice wishes to teleport
to Bob. The field state incident on Alice's beam splitter is
$|\Psi\rangle_{in}=\frac{1}{\sqrt{2}}(|1\rangle_A|0\rangle_B
+|0\rangle_A|1\rangle_B)(a|1\rangle_C+b|0\rangle_C)$, which upon
rearrangement can be written in the Bell basis as
\begin{eqnarray}
|\Psi\rangle_{in}&=&\frac{1}{2}[|\Psi^{(+)}\rangle(a|1\rangle_B
+b|0\rangle_B)+|\Psi^{(-)}\rangle(a|1\rangle_B-b|0\rangle_B)\\
\mbox{}&+&|\Phi^{(+)}\rangle(a|0\rangle_B+b|1\rangle_B)
+|\Phi^{(-)}\rangle(a|0\rangle_B-b|1\rangle_B)]\nonumber
\label{rearrange}
\end{eqnarray}
where $|\Psi^{(\pm)}\rangle$ and $|\Phi^{(\pm)}\rangle$ are the
Bell states defined by
\begin{eqnarray}
|\Psi^{(\pm)}\rangle=\frac{1}{\sqrt{2}}(|0\rangle_A|1\rangle_C\pm
|1\rangle_A|0\rangle_C)\\
|\Phi^{(\pm)}\rangle=\frac{1}{\sqrt{2}}(|1\rangle_A|1\rangle_C\pm
|0\rangle_A|0\rangle_C)\nonumber \label{bell_eq}
\end{eqnarray}
A straightforward algebra based on the quantum theory of the beam
splitter\cite{OC64_485,PRA33_4033} yields that the output states
corresponding to $|\Psi^{(+)}\rangle$, $|\Psi^{(-)}\rangle$,
$|\Phi^{(+)}\rangle$ and $|\Phi^{(-)}\rangle$ are given
respectively by $|0\rangle_E|1\rangle_F$,
$|1\rangle_E|0\rangle_F$,
$\frac{1}{2}(|0\rangle_E|2\rangle_F-|2\rangle_E|0\rangle_F)
+\frac{1}{\sqrt{2}}|0\rangle_E|0\rangle_F$ and
$\frac{1}{2}(|0\rangle_E|2\rangle_F-|2\rangle_E|0\rangle_F)
-\frac{1}{\sqrt{2}}|0\rangle_E|0\rangle_F$, where subscripts $E$
and $F$ refer to the modes of photon exiting the beam splitter via
the output ports $E$ and $F$, respectively. Thus, a detection of a
single photon by the detector $D_F$ combined with a detection of
no photon by the detector $D_E$ would indicate that the input
state is $|\Psi^{(+)}\rangle$ and that, according to Eq.\ (2), the
state at Bob's station is $a|1\rangle_B+b|0\rangle_B$, exactly the
state that Alice wants to teleport to Bob. In this case, Bob needs
do nothing and teleportation is successfully achieved. A detection
of a single photon by the detector $D_E$ and a detection of no
photon by the detector $D_F$ would mean that the input state is
$|\Psi^{(-)}\rangle$. The corresponding state at Bob's station is
$a|1\rangle_B-b|0\rangle_B$. If Bob is informed of such a Bell
measurement result from Alice through classical communication
channels, he needs to apply a $\pi$ phase shifter which changes
the sign of the state $|1\rangle_B$, and teleportation is then
successfully achieved. The teleportation, however, fails, either
if one of the detectors registers two photons and the other none,
which would mean that the input state is $|1\rangle_A|1\rangle_C$,
or if neither detector registers any photon, which would mean that
the input state is $|0\rangle_A|0\rangle_C$. The probability of
success for our teleportation scheme is thus 50\%, which is the
same as the probability of success for the standard teleportation
method. It has been noted \cite{PRA59_116} that a reliable (100\%
probability of success) teleportation cannot be achieved by linear
operations due to the absence of photon-photon interactions. It
should be noted that the 50\% probability of success for our
scheme is obtained only if the Bell states $|\Psi^{(+)}\rangle$
and $|\Psi^{(-)}\rangle$ are clearly distinguished not only from
each other but also from the states $|1\rangle_A|1\rangle_C$ and
$|0\rangle_A|0\rangle_C$ (or from the Bell states
$|\Phi^{(+)}\rangle$ and $|\Phi^{(-)}\rangle$). This means that
our detectors should be capable of distinguishing a single photon
from two. This is of course not an easy requirement to be met. It
seems, however, that single photon counting in the optical regime
and, in particular, in the high-energy(x-ray, $\gamma$-ray) regime
lies within the reach of the present technology. Our analysis also
assumes that the detectors are of unit quantum efficiency.

The state, $a|1\rangle_C+b|0\rangle_C$, to be teleported in our
teleportation scheme can be generated using the methods proposed
in the past\cite{Pegg,Dakna}. One may also generate the state to
be teleported using a beam splitter, as indicated in the leftmost
part of Fig.\ \ref{tele}. The field state emerging from the beam
splitter of complex reflection and transmission coefficients $r$
and $t$ can be written as
$t|1\rangle_C|0\rangle_D+r|0\rangle_C|1\rangle_D$, where the
subscripts C and D refer to the modes of the transmitted and
reflected waves, respectively. The transmitted wave C is then
directed toward Alice's station for teleportation. Alice therefore
has two entangled waves in the state
$|\Psi\rangle_{in}=\frac{1}{\sqrt{2}}(|1\rangle_A|0\rangle_B
+|0\rangle_A|1\rangle_B)(t|1\rangle_C|0\rangle_D
+r|0\rangle_C|1\rangle_D)$ to  be combined in the beam splitter.
She of course has a control over only the waves A and C. The state
$|\Psi\rangle_{in}$ can be rewritten in the Bell basis as
\begin{eqnarray}
|\Psi\rangle_{in}&=&\frac{1}{2}[|\Psi^{(+)}\rangle
(t|1\rangle_B|0\rangle_D+r|0\rangle_B|1\rangle_D)
+|\Psi^{(-)}\rangle(t|1\rangle_B|0\rangle_D
-r|0\rangle_B|1\rangle_D)\\
\mbox{}&+&|\Phi^{(+)}\rangle(t|0\rangle_B|0\rangle_D
+r|1\rangle_B|1\rangle_D)+|\Phi^{(-)}\rangle(t|0\rangle_B|0\rangle_D
-r|1\rangle_B|1\rangle_D)]\nonumber
\end{eqnarray}
If Alice's Bell measurement yields the state $|\Psi^{(+)}\rangle$,
Bob has a wave B in the entangled state
$t|1\rangle_B|0\rangle_D+r|0\rangle_B|1\rangle_D$. The
teleportation is thus successfully achieved. If Alice's Bell
measurement yields the state $|\Psi^{(-)}\rangle$, Bob needs to
apply a $\pi$ phase shifter, which changes the relative phase of
the state $|1\rangle_B|0\rangle_D$ with respect to the state
$|0\rangle_B|1\rangle_D$ by $\pi$. We therefore see that our
scheme offers a simple way of teleporting an entangled state. That
teleportation works also for entangled states was already pointed
out by Bennett et al.
\cite{PRL70_1895} in their original proposal
for quantum teleportation.

It is easy to confirm that teleportation has indeed been
successfully achieved. As shown in the rightmost part of Fig.\
\ref{tele}, we combine the wave D with the teleported wave B using
a beam splitter that has the same transmission and reflection
coefficients as the beam splitter that created the teleported
entangled state $t|1\rangle_C|0\rangle_D+r|0\rangle_C|1\rangle_D$.
If the teleportation is successful, then the input state to the
beam splitter must be
$t|1\rangle_B|0\rangle_D+r|0\rangle_B|1\rangle_D$. The situation
then is exactly the reverse of the situation that created the
teleported entangled state. Thus, a successful teleportation can
be verified by confirming that the detector $D_G$ detects a single
photon and the detector $D_H$ detects none.

Finally we mention that the teleportation scheme described here
uses essentially the same setup as the scheme proposed by Pegg et
al.\cite{Pegg} to perform optical state truncation. The similarity
of the teleportation process and the truncation process has
already been noted by Pegg et al. Whereas the input state to be
truncated is a superposition of many number states including one
photon state and vacuum, and a successful truncation at one photon
state requires waiting until the two detectors register a total of
one photon, the input state to be teleported is a superposition of
one photon state and vacuum, and teleportation is successful half
of the times when the two detectors ($D_E $ and $D_F $ of Fig.\
\ref{tele} )register a total of one photon.

\section{Bell's inequality}

It was shown in the previous section that single-particle
entanglement can be as useful as two-particle entanglement, as far
as application to quantum teleportation is concerned. Considering
that two-particle entanglement provides an opportunity to test
fundamental principles of quantum mechanics related to EPR paradox
and Bell's theorem, one may wonder whether single-particle
entanglement can offer a similar opportunity. Although up to now
Bell's inequality tests have been performed with entangled photon
pairs\cite{RPP41_1881,PRL49_91}, a proposal for an experiment that
demonstrates nonlocality and a violation of Bell's inequality with
a single photon was made 10 years ago\cite{Tan}. The proposal
stimulated much interest and, at the same time, intensive
debate\cite{Santos}. There is no question that the proposed
experiment demonstrates nonlocality of the system and a violation
of Bell's inequality. It, however, does not seem entirely clear at
least to some of the researchers that the outcome of the
experiment can be attributed solely to an effect associated with a
single photon, because the experiment requires performing a
particle-particle correlation measurement.

Here, for our discussion of nonlocality with a single-particle
entangled state, we concentrate on the type of correlation
measurement that can certainly be attributed to a single photon
effect, i.e., a correlation measurement of the first-order type in
Glauber's sense\cite{PR130_2529}. In fact, the nonlocal behavior
demonstrated in the first-order interference measurement of
Grangier et al.\cite{Grangier} with a Mach-Zehnder interferometer
is undoubtedly a single-photon effect. We elaborate further on
this experiment and show that Bell's inequality, which is violated
by the experimental observation of Grangier et al., can be derived
based on the locality assumption. Our argument below can be
considered as a derivation of a single-particle version of Bell's
inequality\cite{PLIC1_195,Ballentine}. We recall that it was
proven\cite{Gisin} that any pure entangled state of two or more
particles violate Bell's inequality. Our derivation allows one to
extend the proof to an entangled state of a single particle. It
should be noted, however, that the interference pattern observed
by Grangier et al. can be explained by a nonlocal classical wave
theory as well as by the quantum theory. A violation of the
single-particle version of Bell's inequality therefore does not
establish the quantum theory as the only correct theory. Its
significance lies in the fact that it gives a quantitative
confirmation that a system described by a single-particle
entangled state behaves nonlocally.

Consider a Mach-Zehnder interferometer consisting of a pair of
lossless symmetric 50/50 beam splitters, each with a pair of
$-\frac{\pi}{2}$ phase shifters, and a pair of perfect mirrors, as
shown in Fig.\ \ref{bell}. A single photon and vacuum are incident
on the first beam splitter from the input ports I and J,
respectively. The output state is again given by
$\frac{1}{\sqrt{2}}(|1\rangle_A|0\rangle_B+|0\rangle_A|1\rangle_B)$.
The reflected wave A and the transmitted wave B are recombined at
the second beam splitter. Alice and Bob, located somewhere along
the pathway of the reflected wave A and the transmitted wave B,
respectively, are each equipped with a phase shifter. If neither
Alice nor Bob applies a phase shifter, the field state emerging
from the second beam splitter is $|1\rangle_C|0\rangle_D$ and it
is certain that the photon strikes the detector $D_C$. Thus, when
N photons are sent from the input port I in succession, all N
photons arrive at the detector $D_C$ and none at the detector
$D_D$. Suppose now Alice inserts her phase shifter into the beam A
and changes its phase by $\phi_A$. A straightforward calculation
based on the quantum theory of the beam
splitter\cite{OC64_485,PRA33_4033} yields that the output state
emerging from the second beam splitter is (apart from an overall
phase factor) $\cos\frac{\phi_A}{2}|1\rangle_C|0\rangle_D
+i\sin\frac{\phi_A}{2}|0\rangle_C|1\rangle_D$. Thus
$N_A\equiv(\sin^2\frac{\phi_A}{2})N$ photons out of the total $N$
incident photons change their paths and strike the detector $D_D$
as a consequence of Alice's action to change the phase of the beam
A by $\phi_A$. If Bob, not Alice, inserts his phase shifter into
the beam B and changes its phase by $-\phi_B$, the output state
becomes $\cos\frac{\phi_B}{2}|1\rangle_C|0\rangle_D
+i\sin\frac{\phi_B}{2}|0\rangle_C|1\rangle_D$. Thus
$N_B\equiv(\sin^2\frac{\phi_B}{2})N$ photons out of the total $N$
incident photons change their paths and strike the detector $D_D$
as a consequence of Bob's action. What would happen if both Alice
and Bob use their phase shifters and change the phases of the
beams A and B by $\phi_A$ and $-\phi_B$, respectively? A
straightforward quantum calculation yields that the output state
in this case is $\cos\frac{\phi_A+\phi_B}{2}|1\rangle_C|0\rangle_D
+i\sin\frac{\phi_A+\phi_B}{2}|0\rangle_C|1\rangle_D$, i.e.
$N_{AB}\equiv(\sin^2\frac{\phi_A+\phi_B}{2})N$ photons out of the
$N$ incident photons change their paths and strike the detector
$D_D$.

On the other hand an argument based on the locality assumption
leads to a result contradictory to the above quantum result. In
order to show this, we assume that those photons that do not
change their paths and still arrive at $D_C$, {\it both\/} when
Alice, not Bob, uses her phase shifter, {\it and\/} when Bob, not
Alice, uses his phase shifter, will still not change their paths
and still arrive at $D_C$ when both Alice  and Bob use their phase
shifters. This assumption means that we do not allow for any
cooperative effect between Alice's phase shifter and Bob's and
therefore assures independence from each other\cite{COMM}. It may
therefore be considered as a single-particle version of the
locality assumption. Let the groups $G_N , G_A , G_B , G_{AB} $
contain, respectively, the total $N$ photons, $N_A$ photons that
strike the detector $D_D$ when Alice, not Bob, uses her phase
shifter, $N_B$ photons that strike the detector $D_D$ when Bob,
not Alice, uses his phase shifter, and $N_{AB}$ photons that
strike the detector $D_D$ when both Alice and Bob use their phase
shifters. The locality assumption dictates that the group $(G_N
-G_A ) \cap (G_N -G_B)$ is a subset of the group $(G_N -G_{AB} )$.
Since the number of photons that belong to the group $(G_N -G_A )
\cap (G_N -G_B )$ is greater than or equal to $N- N_A -N_B $, it
immediately follows that $N-N_{AB} \geq N- N_A -N_B $. We
therefore arrive at the inequality $N_{AB} \leq N_A +N_B $. This
inequality is in disagreement with the quantum theory, because the
inequality, $\sin^2\frac{\phi_A+\phi_B}{2} \leq
\sin^2\frac{\phi_A}{2}+\sin^2\frac{\phi_B}{2}$, is clearly
violated for some values of $\phi_A$ and $\phi_B$. The inequality,
$\sin^2\frac{\phi_A+\phi_B}{2} \leq \sin^2\frac{\phi_A}{2}
+\sin^2\frac{\phi_B}{2}$, is completely equivalent to the formula,
$1+P(\vec{b},\vec{c}) \geq
|P(\vec{a},\vec{b})-P(\vec{a},\vec{c})|$, derived originally by
Bell\cite{PLIC1_195} for a correlated spin pair, if we take the
spin correlation function $P(\vec{a},\vec{c})=-\cos\phi_A $,
$P(\vec{b},\vec{c})=-\cos \phi_B $, and
$P(\vec{a},\vec{b})=-\cos(\phi_A +\phi_B )$.

\section{Conclusion}

In conclusion we have investigated a possibility of utilizing
single-particle entanglement and shown that single-particle
entanglement can be used as a useful resource for fundamental
studies in quantum mechanics and for applications in quantum
teleportation. An experimental scheme that utilizes
single-particle entanglement generally requires production,
maintenance and detection of photons at a single photon level.
With the development of photon counting techniques and of reliable
single photon sources\cite{PRL83_2722}, however, the experimental
realization of the schemes seems within the reach of the present
technology.

\section*{Acknowledgment}

This research was supported by the Brain Korea 21 Project of the
Korean Ministry of Education and by the Korea Atomic Energy
Research Institute(KAERI). The authors wish to thank Professors K.
An, P. Ko, E.K. Lee, S.C. Lee, Y.H. Lee, E. Stewart and Mr. J.C.
Hong for helpful discussions.

\newpage

\begin{figure}
\includegraphics[width=0.8\columnwidth]{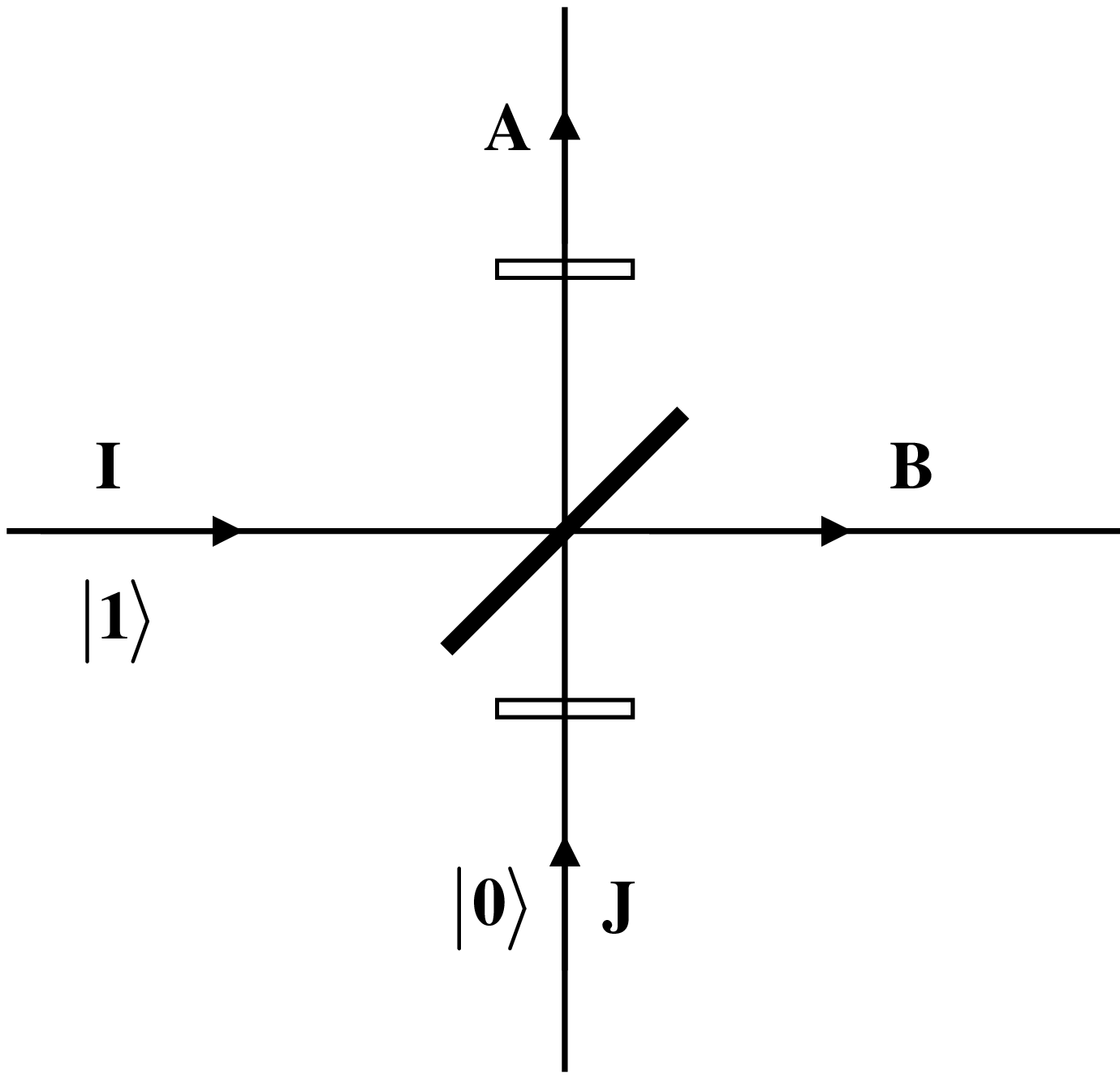}
\vspace{0.3cm}
\caption{\label{beam}}
\end{figure}

\begin{figure}
\includegraphics[width=0.8\columnwidth]{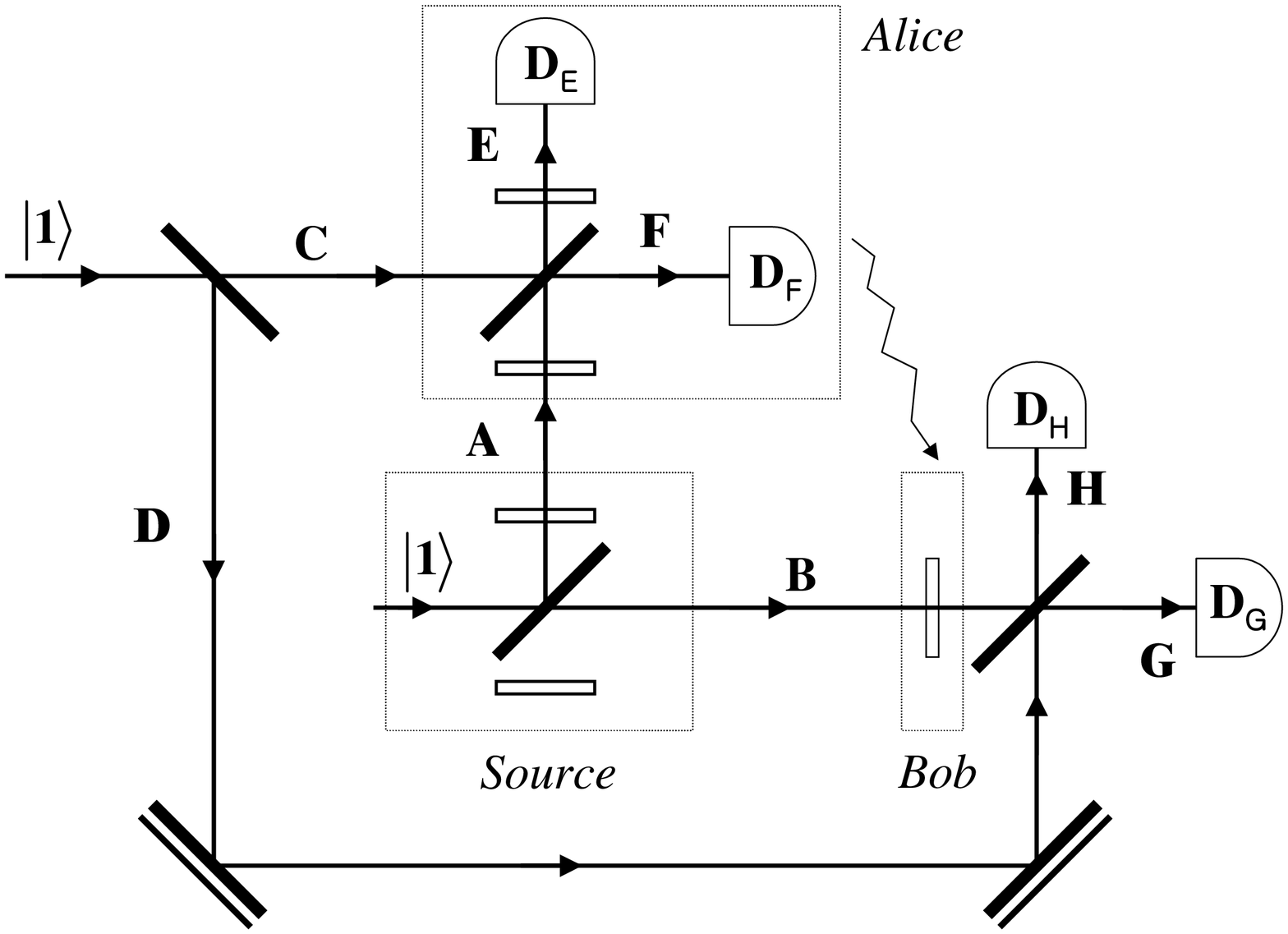}
\vspace{0.3cm}
\caption{\label{tele}}
\end{figure}

\begin{figure}
\includegraphics[width=0.6\columnwidth]{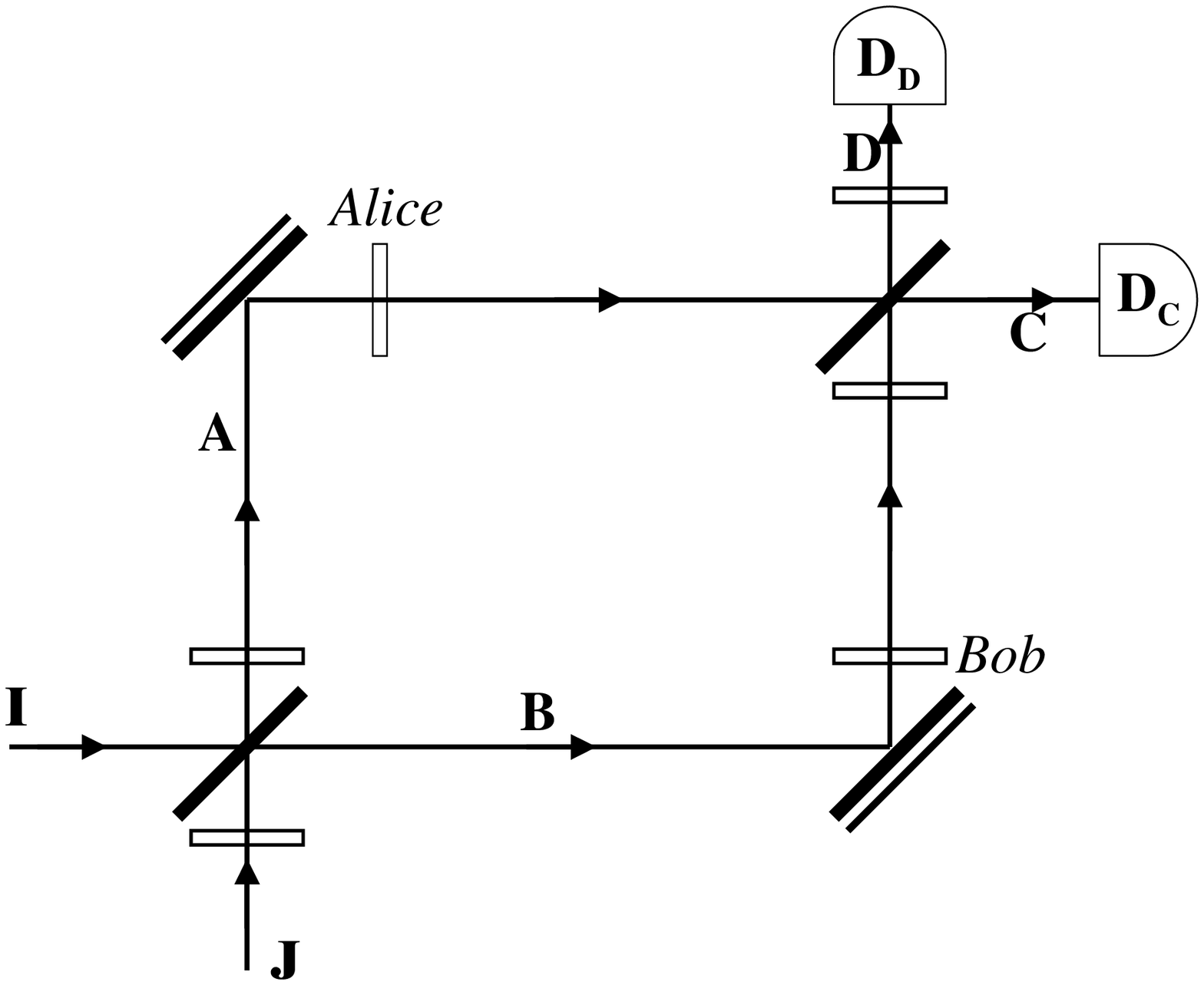}
\vspace{0.3cm}
\caption{\label{bell}}
\end{figure}

\begin{flushleft}

{\Large \bf Figure Captions}

\vspace{\baselineskip}

{\bf Figure \ref{beam}.} Generation of a single photon entangled
state. A single photon and vacuum are incident on a beam splitter
from the input ports I and J, respectively. A $-\frac{\pi}{2}$
phase shifter is placed at the output port A and another at the
input port J.

{\bf Figure \ref{tele}.} Quantum teleportation experiment using
single-particle entanglement. At the source station a single
photon entangled state is generated by a beam splitter. The
transmitted wave B is sent to Bob, while the reflected wave A is
sent to Alice who combines it with the wave C to be teleported.
Alice makes a Bell measurement upon the combined waves A and C and
informs the result to Bob via a classical communication channel
(represented by a wavy line). When Bob is informed of Alice's
measurement result, he performs a suitable unitary transformation
with a $\pi$ phase shifter. The station to the right of Bob
equipped with a beam splitter and detectors $D_G$ and $D_H$
performs a verification of a successful operation of teleportation
if necessary.

{\bf Fig. \ref{bell}.} Single-particle version of Bell's
inequality test with a Mach-Zehnder interferometer. A single
photon and vacuum are incident on the beam splitter (with a pair
of $-\frac{\pi}{2}$ phase shifter) from the input ports I and J,
respectively. The reflected wave A and the transmitted wave B are
recombined at the second beam splitter (with a pair of
$-\frac{\pi}{2}$ phase shifter). Alice and Bob, located somewhere
along the pathway of the reflected wave A and the transmitted wave
B, respectively, each have a phase shifter which they may or may
not use.
\end{flushleft}

\vspace{0.3cm}

\vspace{0.3cm}

\end{document}